\def\href#1#2{{#2}}
\documentclass[aps,pre,floats,amssymb,twocolumn,letterpaper,groupedaddress,
showpacs,floatfix,twoside,tightenlines,superscriptaddress]{revtex4}
\usepackage{graphicx}
\usepackage{amssymb,amsmath}
\usepackage{epsfig}
\begin{document}
\title{Beyond the Wigner Distribution: Schr{\"o}dinger Equations and 
       Terrace Width Distributions}
\author{Howard L.\ Richards} 
\email[Corresponding~author: ]{Howard_Richards@tamu-commerce.edu}
\homepage[]{http://faculty.tamu-commerce.edu/HRichards/index.html}
\affiliation{Department of Physics, University of Maryland, College Park, MD 20742-4111}
\affiliation{Department of Physics, 
     Texas A \& M University--Commerce, 
     Commerce, TX  75429-3011} 
\author{T.~L.\ Einstein}
\email[]{einstein@umd.edu}
\homepage[]{http://www2.physics.umd.edu/~einstein/}
\affiliation{Department of Physics, University of Maryland, College Park, MD 20742-4111}
\date{April 7, 2005} 
\begin{abstract}
The so-called generalized Wigner distribution has earlier been shown to 
be an excellent approximation for the terrace width distribution 
(TWD) of vicinal surfaces characterized by step-step interactions 
that are perpendicular to the average step direction and 
fall off as the inverse square of the step spacing.  In this paper we 
show that the generalized Wigner distribution 
can be derived from a plausible, phenomenological model in which 
two steps interact with each other directly and with other steps through 
a position-dependent pressure. 
We also discuss generalizations to more general step-step interactions 
and show that the predictions are in good agreement with TWDs 
derived from numerical transfer-matrix calculations and 
Monte Carlo simulations.  This phenomenological approach 
allows the step-step interaction to be extracted from experimental 
TWDs. 
\end{abstract}     
\pacs{
  05.40.-a 
  68.35.Md 
  05.70.Np, 
 02.50.-r, 
}

\maketitle

\section{Introduction}

Vicinal surfaces consist of terraces divided by steps 
which interact with each other via a variety of mechanisms, 
including elastic, dipolar, and indirect electronic interactions.  
Since these interactions 
directly determine the distribution of terrace widths, the 
use of terrace width distributions (TWDs) to determine the step-step 
interactions is an important goal.  Theoretical attempts towards 
this goal have 
produced differing results depending upon specific 
approximations \cite{Gruber67,Bartelt90,PM98,IMP98,Masson94,Barbier96,LeGoff99}.

\begin{figure}[t]
\epsfxsize=5.7cm 
\begin{center}
\centerline{\epsfbox{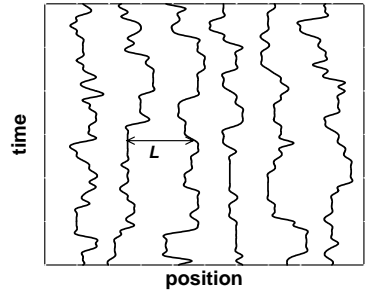}}
\end{center}
\caption[shrt]{
Steps on a vicinal surface can be mapped to the worldlines of 
spinless fermions or hard-core bosons in one spatial dimension (the $x$-direction). 
The ``time'' axis corresponds to the $y$-axis on the vicinal 
surface.
}
\protect\label{fig:Smoothedsteps}
\end{figure}

In many cases, the potential energy $V$ due to the 
interaction between neighboring steps can be written in terms of the 
distance $L$ between the steps as 
\begin{equation}
 \label{eq:InvSquareInteraction}
 V(L) = \frac{A}{L^2} \, .
\end{equation}
This model of interacting steps can be mapped directly onto 
the Calogero-Sutherland model \cite{Calogero69,Sutherland71} of interacting spinless fermions 
(or hard-core bosons) in one spatial dimension (Fig.~\ref{fig:Smoothedsteps}); 
in this picture, the direction along the steps (the 
$y$-direction in ``Maryland notation'') is interpreted as time, and 
the steps themselves are interpreted as the world-lines of the spinless 
fermions.  
The static properties of this system (such as the TWD) depend on $A$ only 
through the dimensionless interaction strength 
\begin{equation}
  \label{eq:deftildeA}
  \tilde{A} \equiv \frac{\tilde{\beta}A}{(k_{\rm B}T)^2} \, ,
\end{equation}
where $\tilde{\beta}$ is the step stiffness, 
$k_{\rm B}$ is Boltzmann's constant, and $T$ is the 
absolute temperature. 
The TWD can be calculated exactly \cite{Calogero69,Sutherland71} when  
$\tilde{A} \! = \! -1/4$, $0$, or $2$. 

An argument \cite{QuantumTransport} from random matrix theory 
originally due to Wigner 
suggests that these three exact TWDs can be approximated by 
\begin{equation}
  \label{eq:Wigner}
  P(s) = a_\varrho s^{\varrho}\exp(-b_\varrho s^2) \, ,
\end{equation}
where $s \! \equiv \! L/\langle L\rangle$,
the constraint $\langle s \rangle = 1$ gives $b_\varrho$ 
[see Eq.~(\ref{eq:omegaEQbrho})], 
$a_\varrho$ is a normalization constant, 
and the relationship 
between $\tilde{A}$ and $\varrho$ is \cite{Calogero69,Sutherland71} 
\begin{equation}
  \label{eq:tildeAvsRho}
  \tilde{A} = \frac{\varrho}{2}\left(\frac{\varrho}{2}-1\right) \, . 
\end{equation}
These ``Wigner distributions" have proven quite successful and have 
been widely used \cite{MehtaRanMat,Haake,Guhr98} 
in nuclear physics (in which case $s$ is the spacing between energy levels),  
random matrix theory, quantum transport, and universal conductance fluctuations. 
There have been a number of 
attempts \cite{Brody73,Berry84,Hasegawa88,Izrailev89,Casati91,AbulMagd96,Kota99}
{}to interpolate between these three special cases.

The so-called generalized Wigner distribution \cite{EP99} 
is defined by the simple assumption
that Eq.~(\ref{eq:Wigner}) and Eq.~(\ref{eq:tildeAvsRho}) 
remain valid for all values of $\tilde{A}$. (This is only 
reasonable for positive or weakly negative values of $\tilde{A}$ \cite{L96}.)  
The generalized Wigner distribution 
appears to be in better agreement with computer 
simulations of vicinal surfaces~\cite{Einstein01} than its competitors, 
and is in reasonable agreement with many 
experimental TWDs \cite{Giesen00}. 
Furthermore, it has been proved that for rational values of 
$\varrho$, $P(s)\! \propto \! s^\varrho$ as 
$s \! \rightarrow \! 0$ \cite{Forrester92,Haldane94,ZNCHa95}.
However, for arbitrary values of $\tilde{A}$ there is no clear 
relation between random matrix theory and the Calogero-Sutherland 
model. This paper provides a formal and physically intuitive 
justification for Eq.~(\ref{eq:Wigner}) for arbitrary values of 
$\tilde{A}$; the resulting formalism allows us to consider more 
general step-step interactions than 
Eq.~(\ref{eq:InvSquareInteraction}). Some of these results have 
been mentioned briefly elsewhere \cite{Einstein01} but are explained here in 
detail.

The organization of this paper is as follows.
In Sec.~\ref{sec-wigner}, we show that the generalized Wigner 
distribution can be derived from a phenomenological model, which 
reduces, under certain circumstances, to the two-particle Calogero model \cite{Calogero69}.
This leads, by separation of variables, to a one-variable Schr{\"o}dinger 
equation that determines the TWD. 
We extend this treatment to non-trivial 
examples in Sec.~\ref{sec-extentions}. 
The significant process of extracting step-step interactions from 
experimental TWDs is discussed in Sec.~\ref{sec-solve}.
Finally, we conclude in Sec.~\ref{sec-conclusions} 
with some possible extensions of this work. 


\section{Physical Explanation of the Generalized Wigner Distribution}
\label{sec-wigner}

\subsection{Deriving a Schr{\"o}dinger equation from the generalized Wigner Distribution}
\label{ssec-quick}

Given a system of steps with interactions of the form specified by 
Eq.~(\ref{eq:InvSquareInteraction}) and a TWD given by the corresponding 
generalized Wigner distribution [Eq.~(\ref{eq:Wigner})], 
we can define a wave function such that 
$\psi_0^2(s) \! \equiv \! P(s)$:
\begin{equation}
  \psi_0(s) = a_\varrho^{1/2} s^{\varrho/2}\exp(-b_\varrho s^2/2)  \, . 
\end{equation}
Differentiating twice, we find
\begin{eqnarray}
  \frac{{\rm d}^2} {{\rm d}s^2} \psi_0(s) 
      & = &  \left[\frac{\varrho}{2}\left(\frac{\varrho}{2}-1\right) s^{-2} 
             - (\varrho+1)b_\varrho + b_\varrho^2 s^2\right]\psi_0(s) \nonumber \\ 
      & = &  \left[\tilde{A} s^{-2} 
             - (\varrho+1)b_\varrho + b_\varrho^2 s^2\right]\psi_0(s) \, ,
\label{eq:diff2}
\end{eqnarray}
where we have used Eq.~(\ref{eq:tildeAvsRho}).
Eq.~(\ref{eq:diff2})  allows us, in somewhat the same spirit 
as the Gruber-Mullins approximation \cite{Gruber67}, to propose 
the following dimensionless Schr{\"o}dinger equation: 
\begin{equation}
  \label{eq:Schroedinger}
  \frac{{\rm d}^2}{{\rm d}s^2} \psi_n(s)  = 
           [\tilde{U}(s)+\tilde{V}(s)-\tilde{E}_n] \psi(s) \, ,
\end{equation}
where
\begin{equation} 
  \label{eq:WignerV}
  \tilde{V}(s) = \tilde{A} s^{-2} 
\end{equation}
is the explicit step-step interaction potential in dimensionless form 
and 
\begin{equation}
  \label{eq:WignerU}
  \tilde{U}(s) = b_\varrho^2 s^2 
\end{equation}
is a dimensionless 
potential due to other steps not explicitly considered.
(This idea is explained in the next subsection.)
 By inspection we see that 
$\psi_0(s)$ is the ground state wave function (it has only one antinode), 
with an eigenvalue given by 
\begin{equation}
  \tilde{E}_0 = (\varrho+1)b_\varrho\, . 
\label{eq:Eo}
\end{equation}

In comparing different experimental 
TWDs with a range of values of $\langle L \rangle$ but with presumably 
the same step-step interaction $V(L)$, it is often useful to 
rewrite Eq.~(\ref{eq:Schroedinger}) in its 
dimensional form
\begin{equation}
  \label{eq:dimSchroedinger}
  \frac{(k_{\rm B}T)^2}{\tilde{\beta}} 
  \frac{{\rm d}^2}{{\rm d}L^2} \Psi_n(L)  = 
           [U(L)+V(L)-E_n] \Psi_n(L) \, ,
\end{equation}
where $V(L)$  is given by Eq.~(\ref{eq:InvSquareInteraction}), 
\begin{equation}
  U(L) \equiv \frac{(k_{\rm B}T)^2} {\tilde{\beta}\langle L \rangle^{2}}
       \tilde{U}\left(\frac{L}{\langle L \rangle}\right)
\end{equation}
and 
\begin{equation}
  E_n \equiv \frac{(k_{\rm B}T)^2} {\tilde{\beta}\langle L \rangle^2}
       \tilde{E}_n. 
\end{equation}
Throughout this paper we will alternate freely and without comment 
between dimensional and dimensionless representations.

Eq.~(\ref{eq:Schroedinger}) can be solved for all the eigenfunctions 
and eigenvectors \cite{QuantumProblems}, so one can 
conveniently consider perturbations from the purely inverse-square 
interaction given by Eq.~(\ref{eq:InvSquareInteraction}).  
The eigenfunctions are
\begin{equation}
  \psi_n(s) = c_n s^{\varrho/2}\exp\left(-b_\varrho s^2\right)
    {}_1\! F_1\left(-n,\frac{\varrho+1}{2},b_\varrho s^2\right) \, , 
\end{equation}
where $n$ is any nonnegative integer, $c_n$ is a normalization 
constant, and 
${}_1\! F_1(a,b,x)$ is Kummer's confluent hypergeometric 
function \cite{Kummer}.
The corresponding eigenvalues are 
\begin{equation}
    \tilde{E}_n = (\varrho+1 + 4n)b_\varrho\, . 
\end{equation}

\subsection{Deriving the generalized Wigner Distribution from a Schr{\" o}dinger equation}
\label{ssec-careful}

An undesirable element of the previous derivation was that it 
implicitly required that one of the steps be held fixed, 
which introduced an artificial (and unnecessary) difference between steps.  
Furthermore, the quadratic nature of 
$\tilde{U}(s)$ was not well explained.  
In this subsection we explore these issues.  

The Calogero-Sutherland \cite{Calogero69,Sutherland71} model with an infinite number of interacting 
spinless fermions would be an ideal model of the vicinal surface, but as mentioned 
above, it is only integrable for three values of $\tilde{A}$.  
A useful alternative is to consider only two adjacent fermions explicitly and model 
the effects of the other fermions phenomenologically through the pressure 
they exert, yielding the Hamiltonian 
\begin{equation}
  \label{eq:Ham2pa}
    {\cal H} = -\frac{1}{2}
    \left(  \frac{\partial^2}{\partial x_1^2} 
                 + \frac{\partial^2}{\partial x_2^2} \right)
           + \tilde{V}(x_2-x_1) - x_1 {\cal P}_1(x_1) + x_2 {\cal P}_2(x_2) \, . 
\end{equation} 
Here ${\cal P}_i$ is the pressure exerted on fermion~$i$, 
which we allow to be position-dependent because the limited 
correlation length in the $x$-direction leads to an effectively 
finite system size \cite{comment}.
Accordingly, we place the ${\cal V}$ fermions immediately to the 
left of fermion~1 in a ``box'' with a fixed left wall at 
$x \! = \! -{\cal V}$; the right wall is at 
$x_1$, the position of fermion~1. The volume (length) of this box is 
${\cal V}_1 \! = \! {\cal V} \! + \! x_1$.
The pressure ${\cal P}_1$ can be expanded as follows: 
\begin{eqnarray}
  {\cal P}_1(x_1) & = & {\cal P} 
         + x_1 \left.\left(\frac{\partial {\cal P}_1}{\partial x_1}\right)\right|_{x_1 = 0}
         + {\cal O}(x_1^2) \label{eq:varpi1a} \\
  & = & {\cal P} 
         + x_1 \left.\left(\frac{\partial {\cal P}_1}{\partial {\cal V}_1}\right)\right|_{{\cal V}_1 = {\cal V}}
         + {\cal O}(x_1^2)    \label{eq:varpi1b} \\
  & = & {\cal P} - x_1 ({\cal V} \kappa)^{-1} + {\cal O}(x_1^2) \, , 
   \label{eq:varpi1c}
\end{eqnarray}
where ${\cal P} \! \equiv \! {\cal P}_1(0)$ and the compressibility $\kappa$ is given by 
\begin{equation}
    \kappa \equiv 
    - \frac{1}{\cal V}
    \left.\left(\frac{\partial {\cal P}_1}
                         {\partial {\cal V}_1}\right)^{-1}\right|_{{\cal V}_1 = {\cal V}} \, . 
\end{equation}
Likewise, we place the ${\cal V}$ fermions immediately to the right of fermion~2 in a 
``box'' of volume (length) ${\cal V}_2 \! = \! {\cal V} \! - \! x_2$, yielding 
\begin{equation}
  \label{eq:varpi2}
  {\cal P}_2(x_2) = {\cal P} + x_2 ({\cal V} \kappa)^{-1} + {\cal O}(x_22) \, . 
\end{equation}
Combining Eqs.~(\ref{eq:Ham2pa}), (\ref{eq:varpi1c}), and (\ref{eq:varpi2}), we 
obtain 
\begin{eqnarray}
    {\cal H} & = &  
        -\frac{1}{2}\left(    \frac{\partial^2}{\partial x_1^2} 
                 + \frac{\partial^2}{\partial x_2^2} \right)
           + \tilde{V}(x_2-x_1) 
    + (x_2-x_1) {\cal P}
\nonumber \\ & & \mbox{}
                 + (x_1^2 + x_2^2) ({\cal V} \kappa)^{-1}
                 + {\cal O}(x_2^3 - x_1^3) \, . 
  \label{eq:Ham2pb}
\end{eqnarray} 
Finally, we can make the following change of variables: 
\begin{eqnarray}
    x_{\rm cm} & = & \frac{x_1 + x_2}{2} \\
    s          & = & x_2 - x_1 \geq 0 \, .  
\end{eqnarray}
This allows us to rewrite the Hamiltonian as follows: 
\begin{eqnarray}
    {\cal H} & = & 
        -\left(  \frac{1}{4}\frac{\partial^2}{\partial x_{\rm cm}^2} 
                 + \frac{\partial^2}{\partial s^2} \right)
           + \tilde{V}(s) 
\nonumber \\ & & \mbox{}
    + [{\cal P} s + ({\cal V} \kappa)^{-1} s^2]
                 + 4({\cal V} \kappa)^{-1} x_{\rm cm}^2 + {\cal O}(sx) \, . 
  \label{eq:Ham2pc}
\end{eqnarray} 
If terms of quadratic and higher order in Eqs.~(\ref{eq:varpi1c}) and (\ref{eq:varpi2})
can be neglected, the Hamiltonian can be separated into a part that 
depends only on $s$ and a part that depends only on $x_{\rm cm}$:
\begin{equation}
  \label{eq:Ham2pd}
    {\cal H} = {\cal H}_{\rm cm} + {\cal H}_s \, , 
\end{equation}
where 
\begin{equation}
  \label{eq:Hamcm}
    {\cal H}_{\rm cm} = -\frac{1}{4}\frac{\partial^2}{\partial x_{\rm cm}^2} 
                 + 4({\cal V} \kappa)^{-1} x_{\rm cm}^2
\end{equation}
and 
\begin{equation}
  \label{eq:Hams}
    {\cal H}_s = -\frac{\partial^2}{\partial s^2}+ \tilde{V}(s) + \tilde{U}(s)
\end{equation}
with 
\begin{equation}
  \label{eq:U}
    \tilde{U}(s) = {\cal P} s + ({\cal V} \kappa)^{-1} s^2 \, . 
\end{equation}

None of the argument so far requires the assumption that  
$\tilde{V}(s)$ is given by Eq.~(\ref{eq:WignerV}).  In the next section 
we will examine how well this phenomenological formalism works 
when applied to other forms of $\tilde{V}(s)$.

If $({\cal V} \kappa)^{-1} \! \gg \! {\cal P}$, we can neglect the term 
${\cal P} s$ in $\tilde{U}(s)$.  The resulting Hamiltonian ${\cal H}$ can 
easily be seen to be the Hamiltonian of the two-fermion Calogero model \cite{Calogero69},
\begin{equation}
  \label{eq:Calogero2part}
  {\cal H} = 
        -\left(    \frac{\partial^2}{\partial x_1^2} 
                 + \frac{\partial^2}{\partial x_2^2} \right)
           + \tilde{V}(x_2-x_1) 
                 + (x_1^2 + x_2^2) \omega^2 \, , 
\end{equation}  
where we identify from Eq.~(\ref{eq:WignerU})  
\begin{equation}
    \label{eq:omega2}
    \omega^2 = ({\cal V} \kappa)^{-1} \, . 
\end{equation}
More importantly, with the identification 
\begin{equation}
  \label{eq:omegaEQbrho}
  \omega = b_\varrho \equiv \left[\frac{\Gamma\left(\frac{\varrho + 2}{2}\right)}
                                       {\Gamma\left(\frac{\varrho + 1}{2}\right)}\right]^2 \, , 
\end{equation}
which again comes from the constraint $\langle s \rangle \! = \! 1$, we 
see that 
\begin{equation}
    \label{eq:HamsSchroedinger}
    ({\cal H}_s - \tilde{E}_n) \psi_n(s) = 0 
\end{equation}
is just Eq.~(\ref{eq:Schroedinger}) and again leads to the generalized 
Wigner distribution. 

Finally, it should be pointed out that the one particle ``Calogero model," 
with a Hamiltonian given by 
\begin{equation}
  \label{eq:GaussHam}
  {\cal H} = 
        -\frac{{\rm d}^2}{{\rm d} x^2} 
                 + x^2  \omega^2 \, , 
\end{equation}  
is really what in essence is used in the many Gaussian 
approximations \cite{Gruber67,Bartelt90,PM98,IMP98,Masson94,Barbier96,LeGoff99}.  These approximations, however, produce conflicting functional relationships 
between $\omega$ and $\tilde{A}$.
This is hardly surprising, since $\tilde{V}$ does not appear explicitly 
in Eq.~(\ref{eq:GaussHam}).  
By using a two-particle Calogero model, we are able to 
state unambiguously the relationship between $\omega$ and $\tilde{A}$. 


\section{Nontrivial Extensions}
\label{sec-extentions}

\begin{figure}[t]
\epsfxsize=5.7cm 
\begin{center}
\centerline{\epsfbox{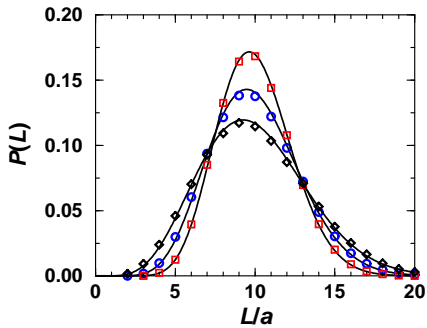}}
\end{center}
\caption[shrt]{
[Color online] TWDs for systems of steps with step-step interactions of the 
form $\tilde{V}(s) \! = \! \tilde{A}_3 s^{-3}$.  
Diamonds, circles, and squares mark TWDs from Monte Carlo 
simulations with $\tilde{A}_3 \! = \! 0.5$, 2, and 6, respectively. 
Solid lines are solutions of Eq.~(\protect\ref{eq:Schroedinger}) 
for the same values of $\tilde{A}_3$ and with 
$\tilde{U}(s)\! \propto \! s^2$.  
}
\protect\label{fig:cubic}
\end{figure}

Here and in the remainder of the text, for our numerical work we use models that are discrete 
in both the $y$-direction and the $x$-direction \cite{Bartelt90,Bartelt92,Masson94,LeGoff99}, in contrast to the continuum step mode, which corresponds to the description given above. 
Specifically, in Monte Carlo simulations 
we use the terrace-step-kink (TSK) model, in which the 
kinks can be any integral number of lattice units long. 
Our transfer-matrix calculations use the restricted TSK model, in which the 
kinks can only be of one lattice unit.  
In both models there are exact expressions for the stiffness of 
an isolated step as a function of temperature \cite{Bartelt92}. 
The interaction between steps is given by a specified function $V(L)$, as with 
the continuum step model.  Further details of our numerical work can be found 
in Refs.~\onlinecite{Einstein01,Hailu04}. 

Eq.~(\ref{eq:Schroedinger}) can be 
extended to step-step interactions of forms other than 
Eq.~(\ref{eq:WignerV}).   
In this section we will present a few arbitrary but 
physically motivated step-step 
interactions and show that the phenomenological method of the 
preceding section still yield TWDs in excellent agreement with 
numerical simulations of the TSK model. 
  
As a first example, we consider interactions of the form 
\begin{equation}
  \label{eq:PowerV}
  \tilde{V}(s) = \tilde{A}_3 s^{-3} \; , 
\end{equation}
which would be a plausible subdominant term in a Taylor expansion of 
the step-step interaction in $s^{-1}$. 
Figure~\ref{fig:cubic} shows TWDs obtained by Monte Carlo 
simulations of steps interacting with this potential 
for three different values of 
$\tilde{A}_3$.  Since the Wigner distribution worked so well with 
nearest-neighbor interactions of the form given by Eq.~(\ref{eq:WignerV}), 
we again take $\tilde{U}(s) \! \propto \! s^2$.
The solutions of the Schr{\"o}dinger equation 
are in excellent agreement with the Monte Carlo data. 

\begin{figure}[t]
\epsfxsize=5.7cm 
\begin{center}
\centerline{\epsfbox{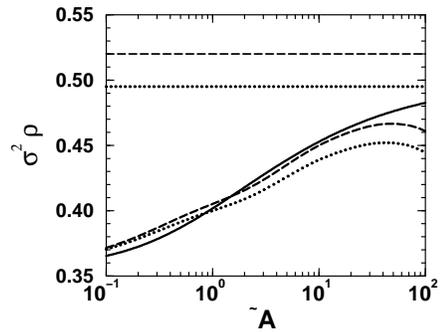}}
\end{center}
\caption[shrt]{
Comparison of $\sigma^2 \varrho$ from the generalized Wigner distribution 
(solid curve) and transfer-matrix calculations for nearest-neighbor-only (NN)
interactions (dashed curve) and interactions out to next-nearest neighbors (NNN)
(dotted curve).  The transfer-matrix calculations were performed with 
5 steps, $\langle L \rangle \! = \! 5$, $\tilde{V}(s)$ given by 
Eq.~(\protect\ref{eq:WignerV}), and $k_{\rm B}T \! = \! 0.84 \epsilon$, where 
$\epsilon$ is the energy of a single kink. 
It appears that the nearest-neighbor-only interactions are in better agreement 
with the generalized Wigner distribution, although this judgment is hampered 
by the small system size and the breakdown, at large $\tilde{A}$, of the 
continuum step model. The dashed and dotted lines indicate limits 
of $\sigma^2 \varrho$ for NN and NNN interactions, respectively, as
derived from Gaussian approximations of the 
the continuum step model in the limit of large $\tilde{A}$ \protect\cite{PM98,IMP98}. 
}
\protect\label{fig:Vrange}
\end{figure}

Interestingly, when {\em all} steps interact (via the potential given 
by Eq.~(\ref{eq:WignerV}), the variance of a TWD with a fixed value of 
$\tilde{A}$ will be smaller than the variance of 
a TWD with the same value of $\tilde{A}$ and just nearest-neighbor steps interacting.  
Numerical evidence supports this 
intuition, and in fact it appears that the Wigner distribution describes 
TWDs from nearest-neighbor interactions better than TWDs from all steps 
interacting (see Fig.~\ref{fig:Vrange}).  On the other hand, the 
phenomenological approach which leads to the generalized Wigner distribution 
depends only on the interaction between nearest neighbor steps, which is 
identical in both cases.  

The difference appears to come from terms in Eq.~(\ref{eq:Ham2pc}) that 
are neglected in Eq.~(\ref{eq:Calogero2part}).  Since the variance is 
decreased by interactions with more steps, it is clear that increasing the 
number of interacting steps will not increase the linear term ${\cal P}$ in 
Eq.~(\ref{eq:Ham2pc}) --- that would increase the variance.  Instead, it appears 
that increasing the number of interacting steps increases the quadratic terms 
in Eqs.~(\ref{eq:varpi1c}) and (\ref{eq:varpi2}).  This would decrease the 
variance and correspond to the intuitive notion that the steps ``interact 
more strongly'' while still observing the constraint $\langle s \rangle \! = \! 1$.
Unfortunately, it also destroys the separability of the Schr{\"o}dinger equation. 

A variety of circumstances, such as the presence of adsorbates \cite{Akutsu} or 
electronic surface states \cite{Pai94,EinsteinPSSS}, can give rise to short-ranged attractive 
forces between steps, oscillating step-step interaction potentials, 
and other complicated interactions.  Under such circumstances, one 
typically finds terraces of two more-or-less well-defined widths;
interactions with steps beyond nearest neighbors can segregate 
these widths, leading to ``step bunching'' \cite{Shenoy}.
This situation is somewhat analogous to liquid-gas coexistence, and 
as with liquid-gas coexistence the compressibility diverges. 
As a result, $\tilde{U}(s)$ is linear in $s$ [Eq.~(\ref{eq:U})]. 

\begin{figure}
\epsfxsize=5.7cm 
\begin{center}
\centerline{\epsfbox{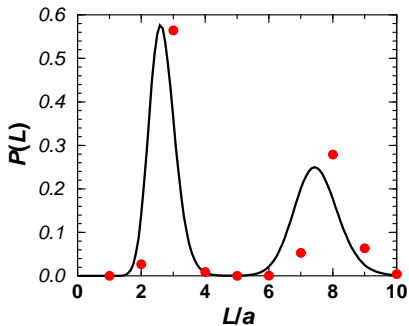}}
\end{center}
\caption[shrt]{
[Color online] A transfer-matrix calculation ($\bullet$) of the TWD for a 
system of five steps with a mean step separation of 5  
lattice units and the oscillating step-step 
potential given by Eq.~(\protect\ref{eq:oscillate}).
The curve is the solution of the corresponding Schr{\"o}dinger 
equation and contains no free parameters.
}
\protect\label{fig:oscillateTWD}
\end{figure}

Vicinal surfaces with both elastic and surface-state mediated 
electronic interactions may be characterized by potentials of the form \cite{Pai94,EinsteinPSSS}
\begin{equation}
  \label{eq:surfstate}
  V(L) =  AL^{-2} + BL^{-3/2} \cos (2k_F L + \phi)
            \, ,
\end{equation}
where $A$ is determined by the elastic interactions, 
$B$ is determined by the coupling to the surface state, 
$k_F$ is the Fermi energy, and $\phi$ is a phase shift. 
The oscillations of this potential can give rise to 
a coexistence as described above, together with a diverging 
compressibility and a linear $U(L)$.  
As a dramatic --- albeit rather unphysical --- example, 
in Fig.~\ref{fig:oscillateTWD}
we compare transfer-matrix TWDs derived from the corrugated interaction potential 
\begin{equation}
  \label{eq:oscillate}
  \tilde{V}(s) = 4 \cos\left(2 \pi s\right)  
\end{equation}
[a special case of Eq.~(\ref{eq:surfstate})] with corresponding 
solutions of Eq.~(\ref{eq:Schroedinger}).
Since the TWDs now have multiple peaks, clearly neither the generalized 
Wigner distribution nor a Gaussian distribution is appropriate.  
Again, considering that there are no free parameters, 
the phenomenological TWD is in excellent 
agreement with the transfer-matrix TWD.  
The slight disagreement is 
probably due to finite-size effects in the transfer-matrix calculation 
(which involved only five steps with $\langle L \rangle \! = \! 5$). 

\section{Solving for the step-step interaction}
\label{sec-solve}

The chief practical interest in terrace width distributions 
comes from a desire to use experimental data to determine the 
interaction potential $V(L)$, or equivalently $\tilde{V}(s)$. 
In this section we 
discuss three progressively more sophisticated methods. 
The first two are appealing in their simplicity, but they suffer from 
serious flaws, and therefore should be avoided. 
By building on them, we introduce the third method, which 
is more computationally intensive than the preceding two, but which has the 
flexibility necessary to be appplied to real experimental data.
For more details see Ref.~\onlinecite{CuTWDs}. 
 

\subsection{Naive direct numerical approach}
\label{ssec-naive}

In Sec.~\ref{sec-wigner},  we derived a Schr{\"o}dinger equation 
that included $\tilde{V}(s)$ in an obvious way, and in 
Sec.~\ref{sec-extentions} we have seen that similar 
Schr{\"o}dinger equations yield TWDs for more complicated 
interactions that are in excellent agreement with numerical simulations of 
the TSK model.  
It is tempting to follow an analogous path when 
dealing with experimental data, using the following procedure: 
\begin{enumerate}
  \item Connect the experimental data points with a smooth interpolating 
    function $P(s)$, such as a cubic spline. 
  \item Calculate $\psi(s) \equiv \sqrt{P(s)}$.
  \item Find ${\rm d}^2\psi/{\rm d}s^2$.
  \item The total potential is 
    $\tilde{V}(s) + \tilde{U(s)} - E \equiv 
    [\psi(s)]^{-1}[{\rm d}^2\psi(s)/{\rm d}s^2]$. 
  \item At large $s$, $\tilde{V}(s)$ is negligible, so 
    the total is given by 
        $\tilde{U}(s) - E$, which should  be constant plus either a 
    quadratic function of $s$ or a linear function of $s$.
  \item Subtract $\tilde{U}(s) - E$ from the potential to 
        recover the step-step interaction $\tilde{V}(s)$.  
\end{enumerate} 

This method has two serious problems.  Experimental TWDs will 
contain measurements indicating $P(s) \! = \! 0$, which lead 
to division by zero in determining the total potential. 
Furthermore, typical statistical fluctuations make the 
numerical estimates of ${\rm d}^2\psi/{\rm d}s^2$ from 
experimental data extremely unreliable. 


\subsection{Fitting the TWD to a preconceived form of $P(s)$}
\label{ssec-preconceivedTWD}

One may eliminate both of these problems by first performing a 
least-squares fit of the TWD to some positive definite and 
twice-differentiable function. 
For example, one might use 
\begin{equation}
  \label{eq:fitP}
 P(s) = c_0 s^{\varrho} (1 + c_1 s + c_2 s^2) \exp (-b s^2) \,  , 
\end{equation}
where $c_0$ is a normalization constant and $c_1$, $c_2$, $\varrho$, 
and $b$ all are parameters to be fitted.   
The fitted $P(s)$ can then be analyzed as above. 

The shortcomings of this second approach are more subtle.  
Eq.~(\ref{eq:fitP}), for example, will always yield a potential that 
diverges as $s^{-2}$ as $s$ approaches zero.  In fact, we have 
fitted Eq.~(\ref{eq:fitP}) to the Monte Carlo data shown in 
Fig.~\ref{fig:cubic}.  Instead of correctly reproducing the 
potential $\tilde{V}(s) \! = \! \tilde{A}_3 s^{-3}$,
the analysis of the fitted $P(s)$ produced an interaction that 
is well approximated by 
$\tilde{V}(s) \! \approx \! \tilde{A}_2 s^{-2}$.
The intersection of this derived potential with the known,  
true potential is for a value of $L$ near $\langle L \rangle$, 
which is not surprising --- the TWD contains more information 
about the interaction where $P(s)$ is large. 


\subsection{Recommended: Fitting the interaction $V(L)$ to a preconceived form}
\label{ssec-preconceivedV}

The method we {\em do} recommend is more computationally demanding, but 
it suffers from none of the above defects: 
\begin{enumerate}
  \item Parameterize $V(L)$, and make a crude initial estimate of the 
        values of the parameters.  
    At least some information on 
    the form of the interaction is often available
        --- whether it is influenced by a surface electronic 
        state, for example.
    In principle, $U(L)$ contains a term linear in $L$ and a 
    term quadratic in $L$, as in Eq.~(\ref{eq:U}); in practice, one can 
    often use either a purely quadratic form of $U(L)$ (as was done to 
    derive the generalized Wigner distribution) or a purely linear form 
    of $U(L)$ (if the compressibility diverges). 
  \item Solve Eq.~(\ref{eq:dimSchroedinger}) numerically for the specified 
        parameters to find $\Psi_0(L)$. 
  \item Find 
        \begin{equation}
             \chi^2 = \sum_{L} \left[\Psi_0^2(L) - P_{\rm exp}(L)\right]^2
        \end{equation}
        for this set of parameters, where $P_{\rm exp}(L)$ is the experimental TWD. 
  \item Iterate procedural steps 2 and 3 in a minimization routine to 
        find parameters for which $\Psi_0^2(L)$ best fits the experimental TWD. 
\end{enumerate}

\begin{figure}[t]
\epsfxsize=5.7cm 
\begin{center}
\centerline{\epsfbox{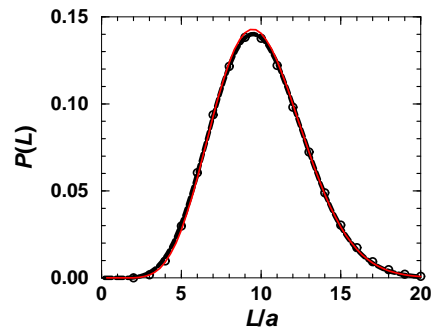}}
\end{center}
\caption[shrt]{
[Color online] Monte Carlo ($\circ$) and phenomenological (solid curve) TWDs for steps interacting with a potential $\tilde{V}(s) \! = \! 2 s^{-3}$.  The dashed curve shows 
a least squares fit using a potential of the form given in 
Eq.~(\protect\ref{eq:fitcubic}). 
}
\protect\label{fig:fitP}
\end{figure}

\begin{figure}[t]
\epsfxsize=5.7cm 
\begin{center}
\centerline{\epsfbox{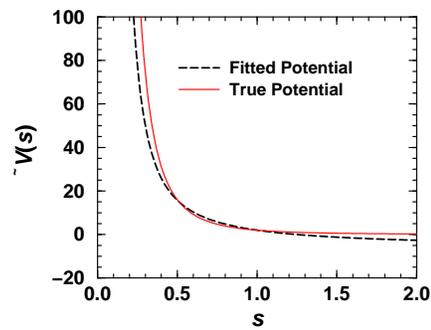}}
\end{center}
\caption[shrt]{       
[Color online] The solid curve shows 
the step-step interaction actually used (solid curve) to 
generate the Monte Carlo TWD shown in Fig.~\protect\ref{fig:fitP}. 
The dashed curve shows the step-step interaction of the form given by 
Eq.~(\protect\ref{eq:fitcubic}) with the parameters given in 
Table~\protect\ref{tab:CubicParameters}, which were determined by a 
least-squares fit of the phenomenological TWD to the Monte Carlo 
TWD.  {}To facilitate a better comparison, the fitted potential has 
been shifted by a constant energy so that it coincides with the 
true potential at $s\! = \! 1$. 
}
\protect\label{fig:fitV}
\end{figure}

\begin{center}
 \begin{table}[b]
 \caption{Parameters in Eq.~(\protect{\ref{eq:fitcubic}}) 
 determined by a least-squares fit of the phenomenological TWD to 
 the Monte Carlo TWD (see Fig.~\protect\ref{fig:fitP}).
 } 
 \label{tab:CubicParameters}
 \begin{tabular}{|l|c|c|c|c|}
     & $\tilde{A}_1$ & $\tilde{A}_2$ & $\tilde{A}_3$ & $\tilde{A}_4$ \\ \hline
 true & 0             & 0             & 2             & 0           \\
 fit  & 6.862        & 1.1035        & 0.40164        & 0.05215       
 \end{tabular}
 \end{table}
 \end{center}

In Fig.~\ref{fig:fitV} we apply this approach to 
the Monte Carlo TWD shown in Fig.~\ref{fig:fitP}.  We parameterize 
$\tilde{V}(s)$ by 
\begin{equation}
  \label{eq:fitcubic}
    \tilde{V}(s) =  \tilde{A}_1 s^{-1} 
                      + \tilde{A}_2 s^{-2} 
                      + \tilde{A}_3 s^{-3} 
                      + \tilde{A}_4 s^{-4} \, ; 
\end{equation}
this form amounts to a generalization of the two-parameter fit 
discussed in Ref.~\onlinecite{statsteps}, where the only allowed term in 
$\tilde{V}(s)$ was $\tilde{A}_2 s^{-2}$. 
The values of the fitted parameters are given in Table~\ref{tab:CubicParameters};
they clearly are quite different from those of the true potential.  
The resulting $\tilde{V}(s)$ is shown in Fig.~\protect\ref{fig:fitV}, 
where it is compared with the actual potential used to generate the 
Monte Carlo data.
What is important, however, is that the fitted and true potentials are close  
over the range in $s$ corresponding to significantly nonzero $P(s)$ --- in 
this case, $0.5 \! \leq  \! s \! \leq 1.5$.  
The phenomenological TWD (Fig.~\ref{fig:fitP}) derived from these fitted parameters is
in good agreement with the Monte Carlo TWD and is very difficult to distinguish 
from the TWD derived from the true potential via Eq.~(\ref{eq:HamsSchroedinger}).  
This result underscores the sensitivity of the process of extracting $\tilde{V}(s)$ and
the necessity of very good statistics in experimental TWDs \cite{statsteps}
and a good idea of the functional form of $\tilde{V}(s)$. 

For this reason, we strongly recommend making a simultaneous fit 
of $V(L)$ to experimental TWDs from {\em several} different misorientations, 
and preferably from {\em several} different temperatures.  We are 
currently applying this approach to Cu surfaces vicinal to the 
(001) plane \cite{CuTWDs}.  We have selected 10 experimental 
TWDs to be fitted, with mean step 
separations ranging from $5.44 a$ to $9.53 a$, where the 
lattice constant $a \! = \! 0.255\mbox{ nm}$, 
and with temperatures ranging from 285~K to 360~K.
Our preliminary studies \cite{Yancey} show that even with experimental data of 
good quality, the extraction of real potentials is not a trivial matter. 

\begin{figure}
\epsfxsize=5.7cm 
\begin{center}
\centerline{\epsfbox{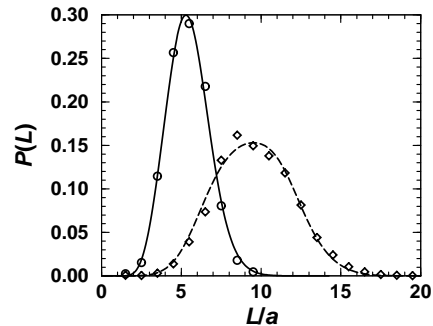}}
\end{center}
\caption[shrt]{
Two experimental TWDs for Cu surfaces vicinal to the (001) plane.
The circles are experimental data courtesy of M.~Giesen, and 
the diamonds are experimental data courtesy of R.~van~Gastel. 
Both experiments were at 295~K.  The corresponding fits
were made by solving Eq.~(\protect\ref{eq:HamsSchroedinger}) 
with the potential shown in Fig.~\protect\ref{fig:Cu.001.V} and 
$U(L) \! \propto \! L$. 
}
\protect\label{fig:Cu.001.TWD}
\end{figure}

\begin{figure}
\epsfxsize=5.7cm 
\begin{center}
\centerline{\epsfbox{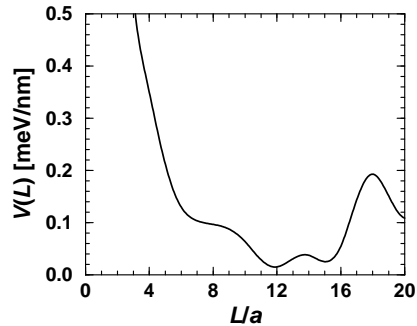}}
\end{center}
\caption[shrt]{
The step-step interaction potential for 
for Cu surfaces vicinal to the (001) plane, 
as determined from fits to ten experimental TWDs 
like those shown in Fig.~\protect\ref{fig:Cu.001.TWD}.
This potential is in good agreement with experimental 
TWDs, but it has probably been ``over-fitted.'' 
}
\protect\label{fig:Cu.001.V}
\end{figure}

Using all 10 experimental TWDs, we were able to obtain good agreement between 
fitted TWDs and experimental TWDs; two examples are shown in Fig.~\ref{fig:Cu.001.TWD}. 
The corresponding interaction potential is shown in Fig.~\ref{fig:Cu.001.V}; 
clearly, it deviates markedly from Eq.~(\ref{eq:InvSquareInteraction}). 
The irregularities in $V(L)$ make it plausible that a coexistence exists between two step 
widths, which would result in a diverging compressibility.  This is supported by the fact 
that the experimental TWDs are fitted with $U(L) \! \propto \! L$ than with 
$U(L) \! \propto \! L^2$; furthermore, the linear form of $U(L)$ would lead to 
``fatter'' tails, as were observed in Ref.~\onlinecite{Giesen00}.

Although these results show internal consistency, the fitted interaction potential 
is markedly different from potentials derived from simulations of surface relaxation 
using an empirical many-body potential \cite{RaouafiSS,Barreteau,Desjonqueres,RaouafiPRB}, 
which are in reasonable agreement with Eq.~(\ref{eq:InvSquareInteraction}).  We believe 
that our interaction potentials are ``over-fitted'' to the experimental data, with   
the irregularities of the $V(L)$ being too strongly influenced by statistical noise 
in our experimental data. This is not really surprising, because $V(L)$ depends quite 
sensitively on $P(L)$.

A similar problem \cite{Robertson93,Mishin99} occurs in determining embedding functions 
for the Embedded Atom Model from experimental data; if too many parameters are fitted, the 
resulting embedding functions of course can fit the supplied experimental data, but 
they give {\em worse} agreement with experimental data not included in the fitting 
process.  The recommended remedy \cite{Robertson93,Mishin99} is to divide the available 
data into two groups, one of which is used to perform the fits, and the other of 
which is used to test the robustness of the fits. 

Presumably, the same remedy will work in our case.  However, a more complete treatment 
of the problem, including estimates of the uncertainties of fitted functions, is 
deferred to a later paper \cite{CuTWDs}.

\section{Conclusions}
\label{sec-conclusions}

We have seen that the generalized Wigner distribution can be 
derived from a  Schr{\"o}dinger equation somewhat in the 
spirit of the Gruber-Mullins approximation, 
and that straightforward extensions of this method work 
for general step-step interactions.  

In Sec.~\ref{sec-wigner} we saw that 
the generalized Wigner distribution can be derived exactly from 
the two-fermion Calogero model \cite{Calogero69}.  This, in turn, can be 
justified from a phenomenological model in which the force on 
two adjacent fermions is derived from the position-dependent pressure 
exerted by other fermions confined in a box, the size of which is 
presumably related to the correlation length in the $x$-direction.  
It is worth noting that the ``entropic repulsion'' is handled implicitly 
by the uncertainty principle in the quantum mapping, so we need 
explicitly consider only the energetic interactions. 

Nothing in this phenomenological picture requires the step-step 
interaction to be given by Eq.~(\ref{eq:WignerV}), and in 
Sec.~\ref{sec-extentions} we demonstrate numerally that 
for very general step-step interactions
the phenomenological picture yields TWDs in excellent agreement with 
numerical simulations of the TSK model. 
This success is particularly impressive when a coexistence between two well-defined 
step widths occurs, as can happen when steps bunch \cite{Shenoy}.  
Under this circumstance 
the compressibility diverges, which causes the tail of the TWD to decay 
exponentially with $L$ rather than exponentially with $L^2$, as it does in 
the case of the generalized Wigner distribution.  This 
phenomenon could explain
the curious slowly decaying tails of TWDs 
mentioned in Ref.~\onlinecite{Giesen00}. 

Since exact solutions are available for both equations in the important  
case in which $\tilde{V}(s)$ is given by Eq.~(\ref{eq:WignerV}), it is 
possible to find not only the TWD but also, using the methods of 
Ref.~\onlinecite{Bartelt92}, an improved estimate of step wandering.
This will be undertaken in a separate paper. 

Finally, although 
we have addressed this work primarily to its surface science applications, 
it may be of interest to research in random matrix theory as well. 
The Calogero-Sutherland model with $\tilde{A}$ = $-1/4$, $0$, or $2$ 
corresponds to random matrices with specific symmetries, and 
attempts to interpolate between them have simply varied the fraction of 
matrices belonging to each symmetry in the ensemble of random 
matrices.  It is far from clear, however, what relation the Calogero-Sutherland model 
with more general values of $\tilde{A}$ or 
our treatment of it has to such mixed ensembles.

\section*{Acknowledgment}

Work was supported by the NSF-MRSEC at U.\ of Maryland
Grant No.\ NSF-DMR 00-80008 
and benefited from interactions with 
J.~D.\ Weeks, M.~Uwaha, and E.~D.\ Williams.  The authors also thank M.~Haftel for 
useful discussions on numerical methods, and M.~Giesen and R.\ van~Gastel for 
experimental TWDs for copper surfaces and helpful comments.  
 

\begin{thebibliography}{10}
\bibitem{comment}
This is a rough argument which is really intended to show only two things: (1)
  the pressure ${\cal P}$ on a fermion is a function of the position of the
  fermion, and (2) the second terms in Eqs.~(\protect{\ref{eq:varpi1c}}) and
  (\protect{\ref{eq:varpi2}}) are proportional to $\kappa^{-1}$. To the extent
  that ${\cal V}$ has a physical meaning, it should be related to the
  correlation length in the $x$-direction, but we have not attempted a careful
  study of this relationship.
\end{thebibliography}


\end{document}